# Title: Random lasers, Lévy statistics and spin glasses: Synergy between photonics and complex systems

**Authors:** Anderson S. L. Gomes[1]*, Ernesto P. Raposo[2], André L. Moura[1,3], Serge I. Fewo[4], Pablo I. R. Pincheira[1], Vladimir Jerez[5], Lauro J. Q. Maia[6], Cid B. de Araújo[1]

**Affiliations:**

[1]Departamento de Física, Universidade Federal de Pernambuco, 50670-901, Recife-PE, Brazil

[2]Laboratório de Física Teórica e Computacional, Departamento de Física, Universidade Federal de Pernambuco, 50670-901, Recife-PE, Brazil

[3]Grupo de Física da Matéria Condensada, Núcleo de Ciências Exatas - NCEx, Campus Arapiraca, Universidade Federal de Alagoas, 57309-005, Arapiraca-AL, Brazil

[4]Laboratory of Mechanics, Department of Physics, University of Yaoundé I, Cameroon

[5]Grupo de investigación FIELDS, Universidad de Investigación y Desarrollo, Bucaramanga, Colombia

[6]Grupo Física de Materiais, Instituto de Física, Universidade Federal de Goiás, 74001-970, Goiânia-GO, Brazil

*Corresponding author. E-mail: anderson@df.ufpe.br

**Abstract**: Random lasers have been recently approached as a photonic platform for disordered complex systems, such as spin glasses. In this work, using a $Nd^{3+}$:$YBO_3$ random laser system operating in the nonresonant (diffusive) feedback regime, we measured the distinct statistics of intensity fluctuations and show the physical origin of the complex interplay between the Lévy regime and the replica-symmetry-breaking transition to the photonic spin-glass phase. A novel result is reported: the unsaturated spin-glass behavior for high excitation pulse energies, not observed for systems with coherent feedback. Our experimental findings are corroborated by the present theoretical analysis. The results herein presented universalize the recent observation consistent with replica symmetry breaking in random lasers with coherent feedback, and also advance on the characterization of the fluctuation statistics of the photonic spin-glass phase, supported by recent theoretical works on the nonlinear optics of complex photonic systems.

**Main Text:** Random lasers (RLs) (*1-3*) have been recently exploited as a photonic platform for theoretical and experimental studies in complex systems, such as spin glasses (SGs) (*4-7*). This multidisciplinary approach opened up new important avenues for the understanding of RL behavior, including Lévy-type distributions of intensity fluctuations (*8-13*) and replica-symmetry-breaking (RSB) phase transitions to a photonic SG behavior (*5-7*), whose experimental evidence was reported for the first time in any physical system using RLs (*14*).

RLs constitute a special class of lasers (*1-3*). They consist of an open cavity system with a disordered gain medium which provides optical amplification, and random scatterers responsible for the necessary feedback. Letokhov proposed the existence of RLs in the late 1960's (*15*), but their first unambiguous observation was only described in 1994 (*16*). Over the last two decades, studies in RLs have grown fantastically, including investigations based on a myriad of cross-disciplinary systems, with examples from biomaterials (*17*) to cancer diagnostic (*18*) and cold atoms (*19*).

RL emission occurs in two regimes regarding the feedback mechanism. In the incoherent or nonresonant regime, the transport mean-free-path of photons, $\ell_t$, is larger than the typical emission wavelength, $\lambda_{em}$, leading to diffusive behavior. Conversely, the coherent or resonant feedback is characterized by values of $\ell_t$ close to the light localization regime, $\ell_t \approx \lambda_{em}$. In the incoherent case, the RL emission spectrum is smooth and narrows around the peak of the gain spectrum. The modes are not explicitly signaled, as described in (*20*). In the coherent regime, the RL spectrum presents spikes that are irrefutable characteristics of resonant feedback loops within the scattering medium, revealing the modes of the system (*21*).

In the present work, we demonstrate the synergy between the RSB SG behavior of RLs and the Lévy statistics of intensity fluctuations when crossing the threshold in a RL system operating in the incoherent (diffusive) feedback regime. In Ref. (*14*) a RSB transition to the SG phase was first described in a RL with *coherent* feedback, but an attempt to observe SG behavior in the incoherent regime did *not* succeed. Here we report for the first time on the SG phase in a RL operating in the *incoherent* regime, showing that the RSB SG transition is universal for RLs.

We also measure the statistical regimes of intensity emission, and report on a remarkable match between the Lévy regime and the critical region of the RSB glassy transition. We further identify the subsequent suppression of the SG behavior in the high-energy Gaussian regime, not reported for RLs presenting coherent feedback. Our work thus advances on the characterization of the fluctuation statistics of the RSB SG phase, which has been comprehensively described in recent theoretical articles (*5-7*) relating nonlinear optics and complex photonic systems.

The RL system investigated here consisted of crystalline powders of $Nd^{3+}$ doped $YBO_3$ (Nd:YBO) – $Nd^{3+}$ concentration: 4.0 mol%. The experiments were conducted with the powder excited by an Optical Parametric Oscillator (OPO) pumped by a Q-switched Nd:YAG laser operating at 806 nm, with 7 ns pulse duration and up to 10 Hz repetition rate. Using a coherent backscattering setup, the measured scattering mean-free-path at 532 nm was 6.5 ± 0.7 μm and the value at 1062 nm was larger than 50 μm, implying $\ell_t > \lambda_{em}$ (see Supplementary Materials).

Figure 1A displays the RL spectral emission of the Nd:YBO system for excitation pulse energies below ($E_p = 1.20$ mJ) and above ($E_p = 1.75$ mJ) the lasing threshold, with the latter showing the smooth signature of the incoherent regime (*1-3*). The RL emission intensity and bandwidth narrowing are shown in Fig. 1B. With basis on the input-output measurement the estimated threshold is $E_{th} = 1.36$ mJ, which closely agrees with the value ($E_{th} = 1.40$ mJ) determined from the full width half maximum (FWHM) of the emitted RL spectrum.

The characterization of the RSB phase transition from the photonic paramagnetic to the SG RL behavior can be quantified (*14*) by the overlap parameter that measures pulse-to-pulse fluctuations in the spectral intensity averaged over $N_s$ laser shots:

$$q_{\gamma\beta} = \frac{\sum_k \Delta_\gamma(k)\Delta_\beta(k)}{\sqrt{\left[\sum_k \Delta_\gamma^2(k)\right]\left[\sum_k \Delta_\beta^2(k)\right]}}, \tag{1}$$

where $\gamma$ and $\beta = 1, 2, \ldots, N_s$, with $N_s = 200$, denote the pulse (replica) labels, the average intensity at the wavelength indexed by $k$ reads $\bar{I}(k) = \sum_{\gamma=1}^{N_s} I_\gamma(k)/N_s$, and the intensity fluctuation is $\Delta_\gamma(k) = I_\gamma(k) - \bar{I}(k)$. The probability density function (PDF) $P(q)$, which describes the distribution of replica overlaps $q = q_{\gamma\beta}$, signalizes a photonic paramagnetic or a RSB SG phase, depending on whether it peaks at $q = 0$ or at values $|q| \neq 0$, respectively.

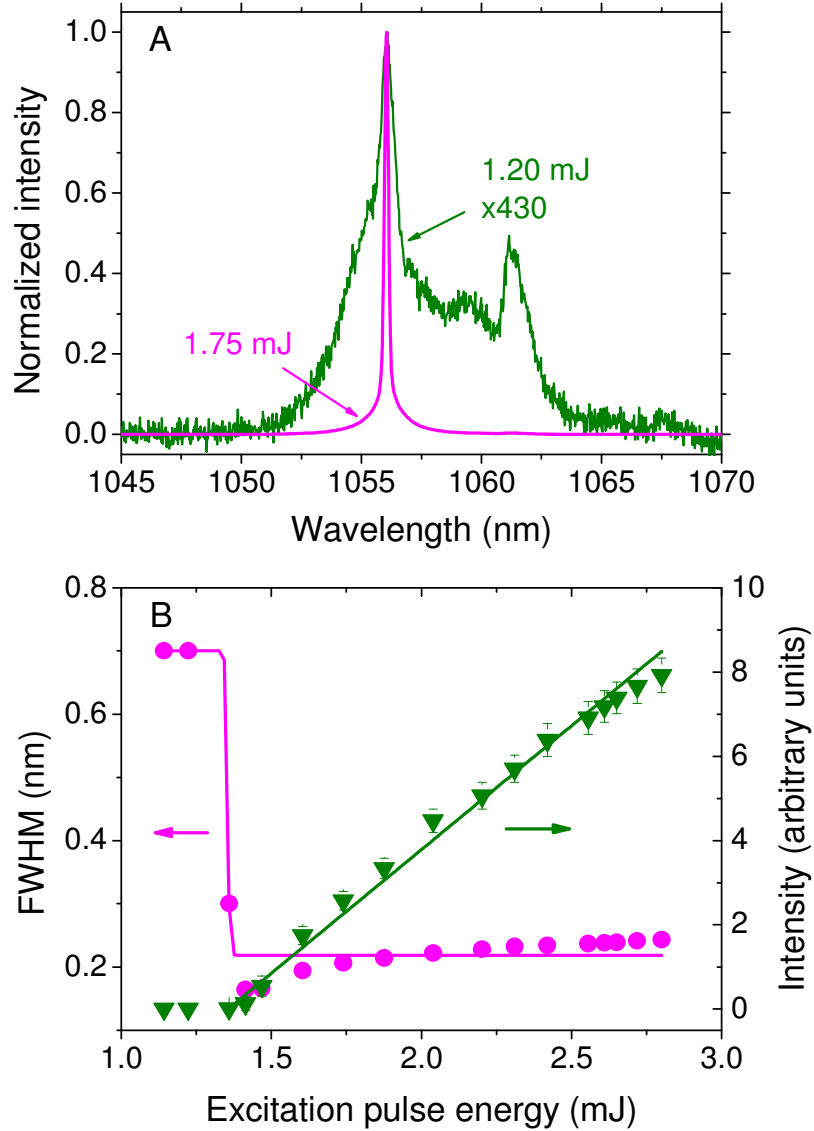

**Fig. 1. Intensity spectra and the RL threshold.** (A) RL spectral emission of the Nd:YBO system in the incoherent regime for two excitation pulse energies: below (green, $E_p = 1.20$ mJ) and above (magenta, $E_p = 1.75$ mJ) the threshold. (B) Emitted intensity (green) and bandwidth narrowing (FWHM, magenta) versus the excitation pulse energy. The intensity measure of the threshold implies $E_{th} = 1.36$ mJ, in close agreement with the FWHM value, $E_{th} = 1.40$ mJ.

In Figs. 2A-2F the pulse-to-pulse intensity fluctuations in the Nd:YBO system can be appreciated, as it evolves from the prelasing (Fig. 2A) to the RL regime (Figs. 2B-2F). The corresponding plots of the PDFs $P(q)$, shown in Figs. 2G-2L, reveal a rich phase structure, emerging from the photonic paramagnetic (Fig. 2G) to the glassy RL behavior above the threshold. For excitation pulse energies below and just above the threshold (Figs. 2G-2I) the photonic behavior is similar to that described in Ref. (*14*) for a RL in the *coherent* regime. However, our results in the *incoherent* feedback regime clearly contrast with those of Ref. (*14*), in which an attempt to observe the SG phase in a colloidal RL system with incoherent feedback was *not* successful. The present report constitutes, therefore, the first demonstration of a photonic RSB SG behavior in a RL operating in the incoherent regime.

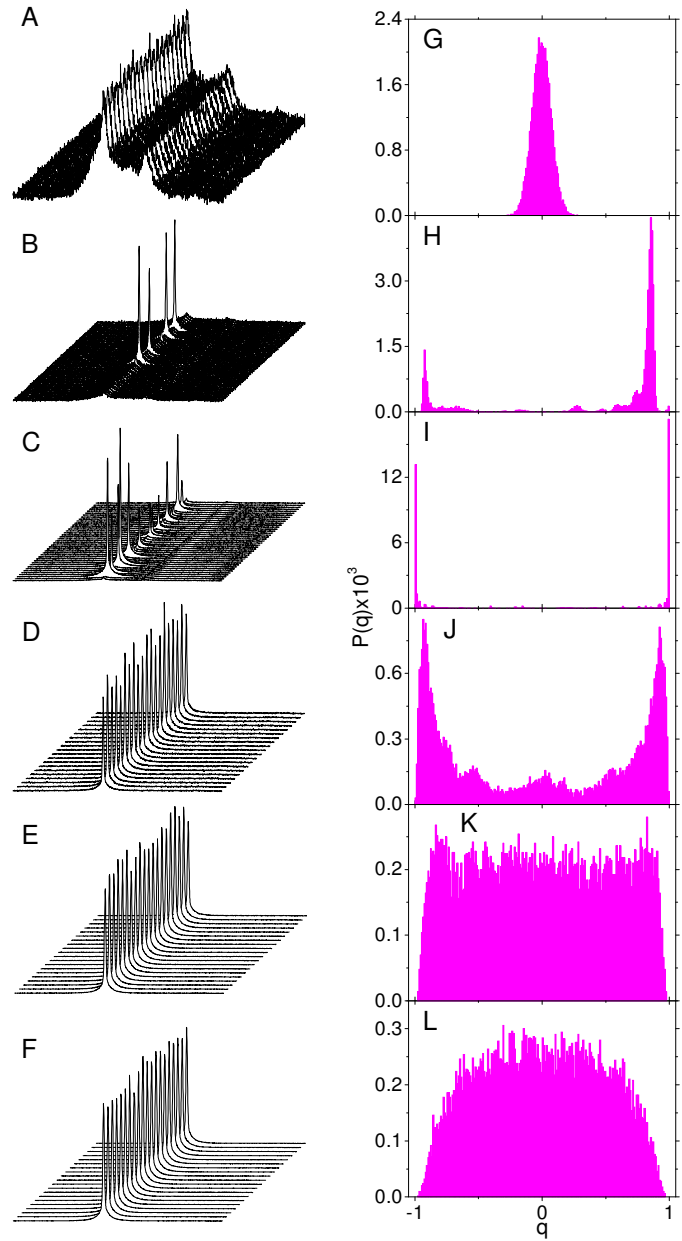

**Fig. 2. Pulse-to-pulse intensity fluctuations and corresponding overlap distributions signalizing the photonic RSB glassy transition in a RL system in the incoherent regime.**
(A)-(F) Intensity spectra showing the pulse-to-pulse fluctuations in the Nd:YBO system in the incoherent feedback regime for excitation pulse energies (A) $E_p = 1.20$ mJ (below the threshold), (B) 1.36 mJ, (C) 1.41 mJ (both around the threshold), (D) 1.60 mJ, (E) 2.20 mJ, and (F) 2.80 mJ (above the threshold). (G)-(L) PDF distributions of the overlap parameter, Eq. 1, corresponding to the data in (A)-(F). Fluctuations are stronger (Lévy-type) close to the threshold, in the critical region of the RSB transition from the photonic paramagnetic to the saturated SG RL behavior. As the excitation pulse energy increases well above the threshold, fluctuations decline considerably (Gaussian regime) and the unsaturated SG behavior emerges, not observed for RLs with coherent feedback (*14*).

For excitation pulse energies well above the threshold, the PDF $P(q)$ starts to broaden (Figs. 2J-2L). This behavior has no counterpart in the coherent regime (*14*), and is related to the deep entry into the Gaussian behavior of intensity emission, as discussed below.

The value $|q| = q_{max}$ at which the distribution $P(q)$ has the maximum defines the so-called Parisi overlap parameter (*4,14*). Its behavior for the Nd:YBO system is displayed in Fig. 3A, indicating the low-energy paramagnetic prelasing ($q_{max} \approx 0$), saturated SG RL ($q_{max} \approx 1$), and high-energy unsaturated SG RL ($q_{max} < 0.9$) regimes. Differently from Ref. (*14*), however, the behavior of $q_{max}$ does not remain nearly constant at the saturation value $q_{max} \approx 1$, but starts to roll over at high energies, consistently with Fig. 2. Well above the threshold the Gaussian fluctuations of intensities decline considerably, decreasing the deviations from the pulse-to-pulse average, $\Delta_\gamma(k)$ and $\Delta_\beta(k)$ in Eq. 1, and leading $q_{max}$ to decay in the high-energy limit of Fig. 3A. This weak-fluctuation regime resembles the behavior of a conventional laser rather than that of a RL, and has not been observed with coherent feedback (*14*).

By analyzing the data in Figs. 2A-2F using the quantile-based McCulloch method (*11,22*), the PDFs of intensities were identified with the family of $\alpha$-stable Lévy distributions, with Lévy index $\alpha \in (0,2]$ and boundary value $\alpha = 2$ corresponding to the Gaussian behavior (*23*). By comparing Figs. 3A and 3B, it is a remarkable fact that, after shifting from the Gaussian ($\alpha = 2$) statistics, the Lévy ($0 < \alpha < 2$) regime corresponds to the narrow critical region of the RSB transition from the photonic paramagnetic ($q_{max} \approx 0$) to the saturated SG RL ($q_{max} \approx 1$) behavior. Actually, due to the sharp bandwidth narrowing around the threshold, the variation in $\alpha$ near the transition is also very acute. Moreover, as the excitation pulse energy increases further a Gaussian RL regime sets in, with $\alpha = 2$, which includes the mentioned unsaturated SG phase.

The complex interplay between the RSB transition to the SG RL phase, and the changes in the statistics of intensity fluctuations can be explained within the same theoretical framework.

Indeed, by considering a disordered nonlinear dielectric medium in a resonant cavity of a RL, the dynamics of the complex amplitudes $a_n(t)$ of the normal electromagnetic modes is driven by the Langevin equations (*5-7*)

$$\frac{da_n}{dt} = -\frac{1}{2}\sum_{pqr} g_{npqr}\, a_q a_r a_p^* + (\gamma_n - \alpha_n)a_n + \eta_n, \qquad (2)$$

where $\gamma_n$ and $\alpha_n$ denote, respectively, the gain and radiation loss coefficient rates, $\eta_n(t)$ accounts for the Gaussian (white) optical noise, and the tensor $g_{npqr}$ marks the nonlinear signature of the spatially disordered medium. The corresponding Langevin equations for the intensities $I_n(t)$ can be derived (*13*), allowing an exact connection (*13,24*) with the associated Fokker-Planck equation. By taking the optical noise as the sum of additive and multiplicative stochastic processes (*13,24*), $\eta_n(t) = \eta_n^{(0)}(t) + a_n(t)\eta_n^{(1)}(t)$, with real correlations $\langle \eta_n^{(1)R}(t)\eta_m^{(1)R}(t')\rangle = Q\delta_{n,m}\delta(t-t')$, the steady-state solution for the PDF of intensities is (*13*) $P(I_n) \propto I_n^{-\mu_n}\exp(-g_{nnnn}^R I_n/2c_n Q)$, with $I_n > 0$ and the power-law exponent $\mu_n = 1 + \sum_{r \neq n}(g_{nrnr}^R + g_{nrrn}^R)/(2Qc_r) - (\gamma_n - \alpha_n)/Q$. As the statistics of the sum of $N$ independent random variables $x$ with power-law distribution $P(x) \propto x^{-\mu}$ is described (*23*) by $\alpha$-stable Lévy distributions with $\alpha = \mu - 1$, the PDF $P(I_n)$ thus identifies an exponentially truncated Lévy distribution of intensities for $0 < \mu_n < 3$ and $0 < \alpha < 2$, with the ultraslow convergence to the Gaussian behavior achieved only for a remarkably large $N$ (*12,23*). For $\mu_n \geq 3$ and $\mu_n \leq 0$ the central limit theorem assures fast convergence to the $\alpha = 2$ Gaussian statistics.

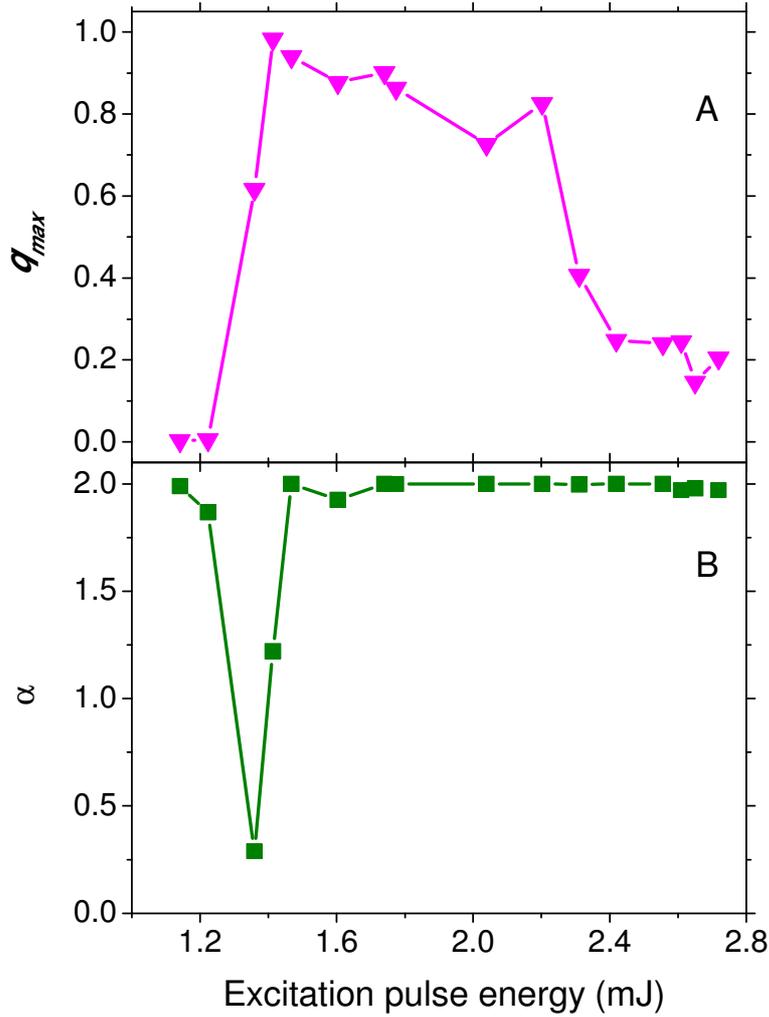

**Fig. 3. Lévy statistics of intensity emission and the RSB glassy transition in a RL system in the incoherent regime.** Dependence on the excitation pulse energy of (A) the Parisi overlap order parameter $q_{max}$ and (B) the Lévy index $\alpha$ calculated from the data in Fig. 2 for the Nd:YBO system in the incoherent feedback regime. The regime of Lévy statistics ($0 < \alpha < 2$) coincides with the critical region of the RSB transition to the SG RL behavior. The value $\alpha = 2$ identifies the Gaussian regimes below and above the transition. Notice that well above the threshold the glassy RL behavior is suppressed, a result not reported for RLs in the coherent feedback regime (*14*).

The connection with the results obtained for the Nd:YBO system can be established by first noticing that, for fixed disorder, the exponent $\mu_n$ increases with the excitation pulse energy $E_p$. For low $E_p$ the intensity signals fluctuate rather weakly in the Gaussian regime, with $\mu_n \leq 0$ (no Lévy-type power-law decay) and $\alpha = 2$ (Figs. 2A and 3B). As the excitation pulse energy enhances, the strong fluctuations shift to Lévy-type behavior, with $0 < \mu_n < 3$ and $0 < \alpha < 2$ (Figs. 2B-2C and 3B). Finally, as $E_p$ increases beyond the threshold fluctuations decline in the Gaussian regime with $\mu_n \geq 3$ and $\alpha = 2$ (Figs. 2D-2F and 3B).

The interplay with the RSB glassy transition lies, on one hand, in the recent proposals (*9,11*) that assign the Lévy index $\alpha$ with the lasing threshold. The threshold value $E_{th} = 1.40$ mJ corresponds to $\alpha \approx 1$, in agreement with (*9*). On the other hand, in a series of remarkable works (*5-7*) a phase diagram for lasers in disordered nonlinear cavities has been recently built also with basis on Eq. 2. In the associated effective Hamiltonian (*5-7*),

$$H = -\frac{1}{2}\sum_{nq} J_{nq}\, a_q a_n^* + \frac{1}{4!}\sum_{npqr} J_{npqr}\, a_q a_r a_p^* a_n^*, \qquad (3)$$

the couplings $J_{nq}$ and $J_{npqr}$ are taken as quenched real Gaussian variables, since the spatial disorder turns the explicit calculation of the tensor $g_{npqr}$ in Eq. 2 rather difficult. The phase diagram displaying the pumping rate versus disorder strength indicates (*5-7*), for large disorder and increasing excitation pump energy, a RSB phase transition at the threshold from the photonic paramagnetic to the SG RL behavior, corroborated by the results presented here for the Nd:YBO system. The Parisi overlap $q_{max}$ acts (*14*) as the order parameter of the RSB transition (Fig. 3A) and, as discussed, the critical region presents Lévy statistics of emission intensities, with saturated SG RL behavior in the Gaussian regime nearly above the transition. At this point, the presence of the unsaturated SG RL regime for high energies emerges as a novel surprising result. Indeed, as the excitation pump energy and the exponent $\mu_n$ increase further above the threshold, the system enters more deeply into the Gaussian regime, leading the fluctuations to weaken considerably (Figs. 2E-2F). Interestingly, this behavior seems to be characteristic of the incoherent regime, in which the spectrum narrows around the peak of the gain transition. Moreover, we can also argue on a possible connection between this loss of intensity overlap and the presence of a new independent disorder term, $-\sum_n \eta_n^{(1)} a_n a_n^*/2$, which has not been previously considered in the Hamiltonian (3). Its origin is related to the multiplicative Gaussian noise in Eq. 2, of null average and standard deviation $\sqrt{Q/2}$, which plays an important role to the emergence of the statistical regimes of intensity fluctuations. Finally, we also remark that the unsaturated SG RL regime has *not* been reported for RLs with coherent feedback (*14*).

In conclusion, our experimental and theoretical results universalize the recent observation of RSB in RLs and also reveal novel unique features of the incoherent (diffusive) feedback regime. The light they shed on the complex interplay between the Lévy statistics of intensity emission and the photonic RSB SG transition can certainly contribute to promote significant advances and unveil new insights in the nonlinear photonics of complex systems.

**References and Notes:**

1. F. Luan, B. Gu, A. S. L. Gomes, K.-T. Yong, S. Wen, P. N. Prasad, *NanoToday* **10**, 168-192 (2015).
2. H. Cao, *J. Phys. A* **38**, 10497-10535 (2005).
3. D. S. Wiersma, *Nat. Photonics* **7**, 188-196 (2013).
4. M. Mézard, G. Parisi, M. A. Virasoro, *Spin Glass Theory and Beyond* (World Scientific, Singapore, 1987).
5. L. Angelani, C. Conti, G. Ruocco, F. Zamponi, *Phys. Rev. Lett.* **96**, 065702/1-4 (2006).
6. F. Antenucci, C. Conti, A. Crisanti, L. Leuzzi, *Phys. Rev. Lett.* **114**, 043901/1-5 (2015).
7. F. Antenucci, A. Crisanti, L. Leuzzi, *Phys. Rev. A* **91**, 053816/1-24 (2015).

8. D. Anglos, A. Stassinopoulos, R. N. Das, G. Zacharakis, M. Psyllaki, R. Jakubiak, R. A. Vaia, E. P. Giannelis, S. H. Anastasiadis, *J. Opt. Soc. Am. B* **21**, 208-213 (2004).

9. S. Lepri, S. Cavalieri, G.-L. Oppo, D. S. Wiersma, *Phys. Rev. A* **75**, 063820/1-7 (2007).

10. R. Uppu, A. K. Tiwari, S. Mujumdar, *Opt. Lett.* **37**, 662-664 (2012).

11. R. Uppu, S. Mujumdar, *Phys. Rev. A* **90**, 025801/1-5 (2014).

12. R. Uppu, S. Mujumdar, *Phys. Rev. Lett.* **114**, 183903/1-5 (2015).

13. E. P. Raposo, A. S. L. Gomes, *Phys. Rev. A* **91**, 043827/1-6 (2015).

14. N. Ghofraniha, I. Viola, F. Di Maria, G. Barbarella, G. Gigli, L. Leuzzi, C. Conti, *Nat. Commun.* **6**, 7058/1-7 (2015).

15. V. Letokhov, *JETP Lett.* **5**, 212-215 (1967).

16. N. M. Lawandy, R. M. Balachandran, A. S. L. Gomes, E. Sauvain, *Nature* **368**, 436-438 (1994).

17. D. Zhang, G. Kostovski, C. Karnutsch, A. Mitchell, *Org. Electron.* **13**, 2342-2345 (2012).

18. R. C. Polson, Z. V. Vardeny, *J. Opt.* **12**, 024010/1-4 (2010).

19. Q. Baudouin, N. Mercadier, V. Guarrera, W. Guerin, R. Kaiser, *Nat. Phys.* **9**, 357-360 (2013).

20. H. E. Türeci, L. Ge, S. Rotter, A. D. Stone, *Science* **320**, 643-646 (2008).

21. H. Cao, Y. G. Zhao, S. T. Ho, E. W. Seelig, Q. H. Wang, *Phys. Rev. Lett.* **82**, 2278-2281 (1999).

22. J. H. McCulloch, *Commun. Stat. Simul.* **15**, 1109-1136 (1986).

23. R. N. Mantegna, H. E. Stanley, *Phys. Rev. Lett.* **73**, 2946-2949 (1994).

24. A. Schenzle, H. Brand, *Phys. Rev. A* **20**, 1628-1647 (1979).

25. R. Corey, M. Kissner, P. Saulnier, *Am. J. Phys.* **63**, 560-564 (1995).

26. G. Labeyrie, C. A. Muller, D. S. Wiersma, Ch. Miniatura, R. Kaiser, *J. Opt. B* **2**, 672-685 (2000).

27. M. A. Noginov, G. Zhu, A. A. Frantz, J. Novak, S. N. Williams, I. Fowlkes, *J. Opt. Soc. Am. B* **21**, 191-200 (2004).

28. M. A. Noginov, *Solid-State Random Lasers* (Springer, New York, 2005).

29. A. L. Moura, S. I. Fewo, M. T. Carvalho, A. N. Kuzmin, P. N. Prasad, A. S. L. Gomes, C. B. de Araújo, *J. Appl. Phys.* **117**, 083102/1-3 (2015).

30. S. García-Revilla, I. Iparraguirre, C. Cascales, J. Azkargorta, R. Balda, M. A. Illarramendi, M. Ai-Saleh, J. Fernández, *J. Opt. Mater.* **34**, 461-464 (2011).

**Acknowledgments:** We thank Robin Kaiser from the Institut Non Linéaire de Nice for enlightening discussions. Work supported by the PRONEX Program, National Institute of Photonics - INFO, CNPq, FACEPE, CAPES, and Finep (Brazilian agencies).

**Supplementary Materials:**

Materials and Methods

Figures S1-S3

References (*25-30*)

Supplementary Materials for

# Random lasers, Lévy statistics and spin glasses: Synergy between photonics and complex systems

Anderson S. L. Gomes, Ernesto P. Raposo, André L. Moura, Serge I. Fewo, Pablo I. R. Pincheira, Vladimir Jerez, Lauro J. Q. Maia, Cid B. de Araújo

**Materials and Methods**

<u>Neodymium crystalline powder: preparation and characterization.</u> The random laser (RL) system used in this work consisted of crystalline powders of $Nd^{3+}$ doped $YBO_3$ (Nd:YBO) obtained by the polymeric precursor method using citric acid ($C_5O_7H_8$, Sigma-Aldrich) as a complexing agent, d-sorbitol ($C_6O_6H_{14}$, Sigma-Aldrich 98%) as a polymerizing agent, and yttrium nitrate hexahydrate ($Y(NO_3)_3 \cdot 6H_2O$, Sigma-Aldrich 99.8%), neodymium hexahydrate ($Nd(NO_3)_3 \cdot 6H_2O$, Sigma-Aldrich 99.8%) and boric acid ($H_3BO_3$, Ecibra 99.5%) as precursors for Y, Nd and B, respectively. The synthesis of the material was achieved by dissolving the yttrium and neodymium nitrates in an aqueous solution of citric acid at room temperature. This solution was added to another solution of d-sorbitol and boric acid previously dissolved in water. The obtained solution was then annealed at 150 °C, whereby the polymerization process occurred, forming a dried resin. The molar ratio of citric acid to elements (metals + boron) was 3:1. The citric acid/d-sorbitol mass ratio was set to 3:2. The dried resin was calcinated at 400 °C during 24 h and heat-treated at 900 °C/1 h.

The X-ray diffraction (XRD) measurements were taken with a Shimadzu XRD-6000 X-ray diffractometer with Bragg-Brentano theta-2 theta geometry, at a continuous scan speed of 1°/min from 5° to 80° with sampling pitch of 0.01°. $K\alpha$ radiation of 1.54059 Å from a Cu tube operating at 40 kV was used.

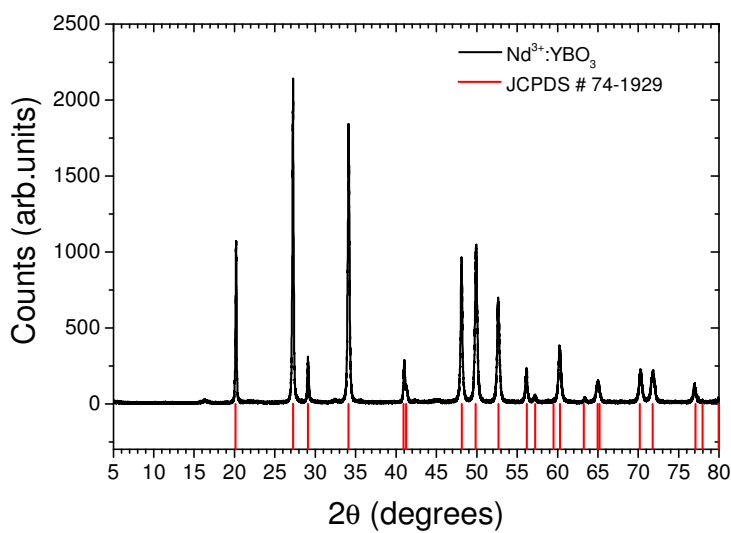

**Figure S1. X-ray diffraction pattern of the 4.0% Nd:YBO crystalline powder.** The pattern was taken from the JCPDS card number 74-1929.

Figure S1 shows the XRD pattern of the sample of Nd:YBO crystalline powder with $Nd^{3+}$ concentration of 4.0 mol%. For comparison, the diffraction positions from the JCPDS card number 74-1929 of $YBO_3$ were also included. The $YBO_3$ phase is an hexagonal structure with $P6_3/mmc$ (194) space group and centro-symmetric.

The samples were characterized structurally using a JEOL JEM 2010 high-resolution transmission electron microscope (HRTEM) operating at 200 keV. Figure S2 shows TEM, HRTEM and SAED (selected area of electron diffraction) images of the sample. The crystalline powders display oval and spherical shaped grains. In fact, the TEM and HRTEM provide a two-dimensional picture of the well-crystallized nanoparticles (Figs. S2C-D) and the coalescence effect is observed (Figs. S2A-B).

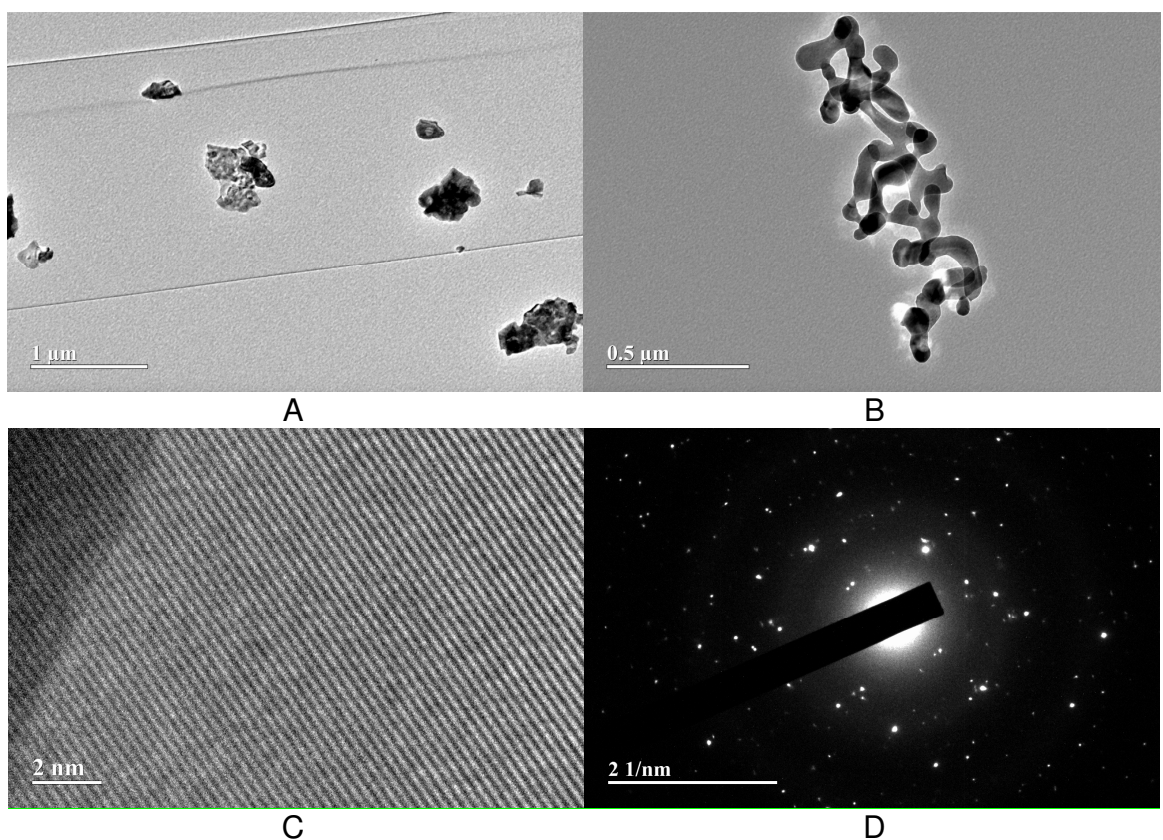

**Figure S2. TEM, HRTEM and SAED images of the 4.0% Nd:YBO crystalline powder.** (A)-(B) TEM images of some particles, (C) HRTEM (high-resolution TEM) image of an individual particle showing the structural planes, and (D) the SAED (selected area of electron diffraction) image.

The sizes distribution obtained by measuring 268 particles is shown in Fig. S3. Particles with sizes ranging from ≈40 to ≈1000 nm were observed, with the highest occurrence lying around 120 nm. The solid curve in Fig. S3 represents the best fit to a log-normal function.

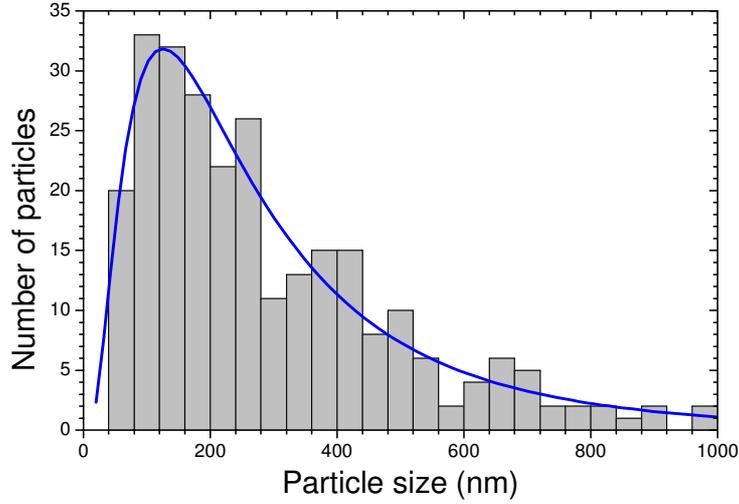

**Figure S3. Sizes distribution histogram of the 4.0% Nd:YBO powder grains.** The highest occurrence of particle sizes lies around 120 nm. The solid curve represents the best fit to a log-normal function.

Determination of the mean-free-path and the RL feedback regime. For the particle size with highest occurrence, the scattering mean-free-path, $\ell_s$, defined as the average distance traversed by a photon between successive scattering events, was measured by a conventional coherent backscattering experiment at 532 nm, using an experimental setup with a cw laser similar to those of Refs. (*25,26*). For the 4.0% Nd:YBO system used in this work, the value $\ell_s$ = 6.5 ± 0.7 μm was inferred. In a previous study (*27*), the dependence on the particle size of the transport mean-free-path, $\ell_t$, which gives the average distance traversed by photons before changing direction, was measured and calculated. For particle sizes around 120 nm, the estimates for $\ell_t$, which is close to $\ell_s$, lie in the ranges 5-10 μm and 50-80 μm, at 532 nm and 1056 nm, respectively. Therefore, in either cases the transport mean-free-path is larger than the typical emission wavelength, $\ell_t > \lambda_{em}$, corroborating the fact that the RL system investigated in this work operates in the incoherent or nonresonant (diffusive) feedback regime.

Optical experiments. The optical experiments were conducted with the powder excited by an Optical Parametric Oscillator (OPO) pumped by the second-harmonic of a Q-switched Nd:YAG laser (7 ns, 10 Hz). The powder was placed on a sample holder and gently pressed into a uniform disc region. The light beam from the OPO was focused on the sample by a 10 cm focal length lens. The illuminated area was 1.8 mm². The excitation wavelength, 806 nm, in resonance with the $^4I_{9/2} \rightarrow {}^4F_{5/2}$ transition, was chosen to optimize the fluorescence signal around 1060 nm due to the $^4F_{3/2} \rightarrow {}^4I_{11/2}$ transition, leading to a RL emission peaking at 1056 nm. We remark that $Nd^{3+}$-doped RLs are well characterized in the literature (*29-30*), but they have been used here for the first time as a platform for complex systems experiments, with the advantage of being solid state, without movement of scatterers as in colloidal systems, and the fact that they also act as the gain media.

Although the results shown in the main text employed the 4.0% $Nd^{3+}$ concentration, we also remark that samples with 0.5%, 1.0%, 1.5% and 2.0% were also studied and, except for the

0.5% and 1.0% concentrations in which RL emission was not observed, the other two samples behaved qualitatively in the same way as the 4.0% sample.

**References**

25. R. Corey, M. Kissner, P. Saulnier, *Am. J. Phys.* **63**, 560-564 (1995).

26. G. Labeyrie, C. A. Muller, D. S. Wiersma, Ch. Miniatura, R. Kaiser, *J. Opt. B* **2**, 672-685 (2000).

27. M. A. Noginov, G. Zhu, A. A. Frantz, J. Novak, S. N. Williams, I. Fowlkes, *J. Opt. Soc. Am. B* **21**, 191-200 (2004).

28. M. A. Noginov, *Solid-State Random Lasers* (Springer, New York, 2005).

29. A. L. Moura, S. I. Fewo, M. T. Carvalho, A. N. Kuzmin, P. N. Prasad, A. S. L. Gomes, C. B. de Araújo, *J. Appl. Phys.* **117**, 083102/1-3 (2015).

30. S. García-Revilla, I. Iparraguirre, C. Cascales, J. Azkargorta, R. Balda, M. A. Illarramendi, M. Ai-Saleh, J. Fernández, *J. Opt. Mater.* **34**, 461-464 (2011).